\documentstyle[prd,aps,epsfig,preprint]{revtex}
\tighten
\begin{document}
\thispagestyle{empty}
\baselineskip24pt
\draft
\begin{center}
{\large\bf Octet Magnetic Moments with Null Instantons and Semibosonized 
Nambu-Jona-Lasinio Model}
\end{center}
 \begin{center}
{Elena~N.~Bukina}
\hbox{\it Laboratory of Theoretical Physics,}
\hbox{\it Joint Institute for Nuclear Research,
          141980 Dubna, Moscow region, Russia} 
\hbox{\it e-mail: bukina@thsun1.jinr.ru,} 
\end{center}
\date{\today}
\begin{abstract}
It is shown that the difference between the magnetic moment
results in the quark model with null instantons and semibosonized
Nambu-Jona-Lasinio model lies in the description of the
magnetic moment of the $ \Lambda$-hyperon.
\end{abstract}


\section{Introduction}
\setcounter{equation}{0}

Main features of the baryon magnetic moments have been understood in the 
frameworks of unitary symmetry \cite{Gl} and quark models \cite{Morp} 
years ago. But nowadays experimental accuracy of the baryon magnetic moments 
measurement is very high reaching the level of $1\%$ \cite{Mont}.
So theoretician task is no more to describe them reasonably well but to
search more elaborated models to account for data. 
Among many interesting models developed in order to solve this problem from
various points of view I cite only few of them 
\cite{Sehg1}-\cite{Kim}. Recently in \cite{Chpi} we have shown 
that the baryon
magnetic moment description in frameworks of the chiral model 
\cite{Chang} and the quark soliton $\xi$QSM one \cite{Kim}
are practically
identical. The main difference from the traditional unitary
symmetry approach proves to be in terms corresponding to some kind of the
meson cloud contribution and in the prediction for the $\Lambda$
hyperon moment.
In this work
I want to investigate the unitary symmetry content of two recent rather
sophisticated models, that of \cite{Iwao} and of
aforementioned \cite{Kim}. 
 
I shall show that the difference between the magnetic moment
results in the quark model with null instantons and semibosonized
Nambu-Jona-Lasinio model lies principally in the evaluation of the
magnetic moment of the $ \Lambda$-hyperon.

\section{Magnetic moments with null instantons}
\setcounter{equation}{0}

First I consider an interesting model \cite{Iwao} with null instantons where
a dynamical symmetry breaking for the magnetic moments of baryons
was proposed. This effect was 
attributed to the contribution arising due to the time average
of the two quark exchange in spin 0 state, contained in an
instanton and anti-instanton ball, into the magnetic moment of
the third quark.
We remind here the main formulae of \cite{Iwao}:

\begin{equation}
\begin{array}{ll}
\mu(p)=(1+\frac{8}{9}z)\mu-\alpha, &
\mu(n)=-\frac{2}{3}\mu(1+\frac{1}{3}z), 
\nonumber\\
\mu(\Sigma^{+})=\frac{1}{9}(8(1+z)\mu+\mu^{'})-\beta, &
\mu(\Sigma^{-})=\frac{1}{9}(-4\mu+\mu^{'})+\beta, 
\nonumber\\
\mu(\Xi^{0})=\frac{1}{9}(-4\mu^{'}-2(1+z)\mu), & 
\mu(\Xi^{-})=\frac{1}{9}(-4\mu^{'}+\mu)+\beta,  
\nonumber\\
\mu(\Lambda)=-\frac{1}{3}\mu^{'}-\frac{1}{3}(\alpha-\beta). 
\end{array}
\label{iw}
\end{equation}

Now let us analyze the unitary symmetry content of the Eqs.~(\ref{iw}).
It is straightforward to see that the SU(3) electromagnetic current is 
no longer a pure octet but is broken with the Gell-Mann - Okubo \cite{GMO,Ok} and other
terms (I disregard space-time indices):
\begin{eqnarray}
J^{e-m,symm}= -F(\overline{B}^{\gamma}_{1}B_{\gamma}^{1}-
\overline{B}^{1}_{\gamma}B_{1}^{\gamma})+
D(\overline{B}^{\gamma}_{1}B_{\gamma}^{1}+
\overline{B}^{1}_{\gamma}B_{1}^{\gamma})+ \nonumber\\
g_{1}\overline{B}_{3}^{\gamma}B^{3}_{\gamma}+
g_{2}\overline{B}_{\gamma}^{3}B^{\gamma}_{3}+(T-
\frac{2}{3}D)Sp(\overline{B}^{\gamma}_{\beta}B_{\gamma}^{\beta})+
3d(\overline{B}_{2}^{1}B_{1}^{2}-\overline{B}_{1}^{2}B_{2}^{1}),
\label{fdiw}
\end{eqnarray}
where
\begin{equation}
\begin{array}{ll}
\mu(p)=F+\frac{1}{3}D+g_{1}+ t,  &  
\mu(n)=-\frac{2}{3}D+g_{1}+ t ,
\nonumber\\
\mu(\Sigma^{+})=F+\frac{1}{3}D+t +d, & 
\mu(\Sigma^{-})=-F+\frac{1}{3}D+t -d, 
\nonumber\\
\mu(\Xi^{0})=-\frac{2}{3}D+g_{2}+t, &
\mu(\Xi^{-})=-F+\frac{1}{3}D+g_{2}+t,
\nonumber\\
\mu(\Lambda)=-\frac{1}{3}D+\frac{2}{3}(g_{1}+g_{2})+t 
\end{array}
\label{fd}
\end{equation}
and 
\begin{equation}
\begin{array}{ll}
D=(\mu_{u}-\mu_{d})-\frac{1}{2}(\alpha-\beta), &
F=\frac{2}{3}(\mu_{u}-\mu_{d})-\frac{1}{2}(\alpha+\beta), 
\nonumber\\
g_{1}=-\frac{1}{3}(\mu_{d}-\mu_{s})-\frac{1}{2}(\alpha-\beta), &
g_{2}=-\frac{5}{3}(\mu_{d}-\mu_{s})-\frac{1}{2}(\alpha-\beta),
\nonumber\\
d=\frac{1}{2}(\alpha-\beta), &
t+\frac{1}{3}D=\frac{2}{3}(\mu_{u}+\mu_{d})-\frac{1}{3}\mu_{s}.
\label{ggg}
\end{array}
\end{equation}
Here \cite{Iwao}
$$ \mu_{u}=\frac{2}{3}(1+z)\mu,\quad \mu_{d}=-\frac{1}{3}\mu, 
\quad \mu_{s}=-\frac{1}{3}\mu^{'}$$
and $B^{\gamma}_{\eta}$ is a baryon octet, $B^{3}_{1}=p$,
$B^{2}_{3}=\Xi^{0}$ etc.
This current is just a sum of the traditional unitary electromagnetic
current \cite{Gl} and the traditional unitary middle-strong baryonic one 
\cite{GMO,Ok} plus a nonlinear in quantum numbers term for  $ \Sigma$
-hyperons. The latter term can be attributed to some kind of meson cloud 
contribution. With some algebra
it can be placed into nucleon magnetic moments or $\Xi$ hyperon ones. 
In the model based on the phenomenological current Eq.(\ref{fdiw}) the only
relation holds (the numbers under this and following relations are
given in nuclear magnetons):

\begin{equation}
2\mu(p)+2\mu(n)-\mu(\Sigma^{+})-\mu(\Sigma^{-})+2\mu(\Xi^{-})+
2\mu(\Xi^{0})-6\mu(\Lambda)=0 
\end{equation}
$$( 10.42-10.04=0).$$ 
As due to Eq.(\ref{ggg})
\begin{equation}
5g_{1}-g_{2}+4d=0,
\label{ggd}
\end{equation}
it splits into two relations Eq.(\ref{rel1}) and Eq.(\ref{rel2}) tacitly 
contained in \cite{Iwao}:

\begin{equation}
10\mu(p)+20\mu(n)-5\mu(\Sigma^{+})-13\mu(\Sigma^{-})+18\mu(\Xi^{-})-
30\mu(\Lambda)=0 
\label{rel1}
\end{equation}
$$( 61.4-62.2=0),$$
\begin{equation}
4[\mu(\Sigma^{-})-\mu(\Xi^{-})]-5[\mu(n)-\mu(\Xi^{0})]=0
\label{rel2}
\end{equation}
$$(-2.04+3.30=0).$$
The most successful fit of \cite{Iwao} (those named (iv)) reads:
$$ \mu=3.02398909,\quad \mu^{'}=1.18108,
\quad z=-0.1532,\quad \alpha=\beta=-0.18065841. $$

So the main features of the baryon magnetic moment picture in the
quark model with instantons of \cite{Iwao} 
can be comprehended in terms of unitary symmetry approach.

\section{Semibosonized Nambu-Jona-Lasinio model for the
baryon magnetic moments}
\setcounter{equation}{0}

Now I consider another quite distinct model for the magnetic moments of 
baryons developed in \cite{Kim}
within the chiral quark soliton model. In this model, known also as
the semibosonized Nambu-Jona-Lasinio model, the baryon can be
considered as $N_{c}$ valence quarks coupled to the polarized Dirac sea
bound by a nontrivial chiral background hedgehog field in the Hartree-Fock
approximation \cite{Kim}. Magnetic moments of baryons were written in the
form \cite{Kim}:

\begin{equation}
\left(\begin{array}{ccccccc}\mu(p)\\ \mu(n)\\ \mu(\Lambda)\\
\mu(\Sigma^{+})\\ \mu(\Sigma^{-})\\ \mu(\Xi^{0})\\ \mu(\Xi^{-})
\end{array}\right)=
\left(\begin{array}{ccccccc}-8&4&-8&-5&-1&0&8\\6&2&14&5&1&2&4\\
3&1&-9&0&0&0&9\\-8&4&-4&-1&1&0&4\\2&-6&14&5&-1&2&4\\
6&2&-4&-1&-1&0&4\\2&-6&-8&-5&1&0&8\end{array}\right)
\left(\begin{array}{ccccccc}v\\w\\x\\y\\z\\p\\q\end{array}\right)
\label{kim}
\end{equation}
Here parameters  $v,w$ are related linearly with the usual 
coupling constants of the unitary symmetry approach \cite{Gl} while parameters
$x,y,z,p,q \simeq m_{s}$ are specific for the model. Upon algebraic
transformations the expressions for 6 baryons $B(qq,q^{'})$
can be rewritten in the form analogous to Eqs.(\ref{fd})
where now

\begin{equation}
\begin{array}{ll}
F=-5v+5w-9x-3y-p-2z, &
D=-9v-3w-13x-7y+4q-p,\nonumber\\
g_{1}=4x+4y-4q-z, &
g_{2}=22x+10y-4q+2p+z,\nonumber\\
t=\frac{1}{3}(28x+13y+8q+4p).
\end{array}
\end{equation}
It means that magnetic moments of the octet baryons B(qq,q') in the 
semibosonized Nambu-Jona-Lasinio model \cite{Kim} can also be obtained 
from the unitary electromagnetic current given by Eq.(\ref{fdiw}).
With this current the magnetic moment of the $\Lambda$-hyperon reads:

\begin{equation}
\mu(\Lambda)^{null}=-\frac{1}{3}D-(8x+5y-8q),
\end{equation}
which differs from that given by Eq.(\ref{kim}):

\begin{equation}
\mu(\Lambda)^{NJL}-\mu(\Lambda)^{null}=-\frac{1}{3}
(16x-8y-7q+p).
\end{equation}

Besides the relation given by Eq.(\ref{ggd}) this is the only important
difference in predictions of the two otherwise independent models. 

\section{Conclusion}

So the two models with different theoretical foundations 
that is the quark model with null instanton and anti-instanton balls \cite{Iwao} 
and semibosonized Nambu-Jona-Lasinio model \cite{Kim} give 
similar algebraic structure for the 
magnetic moments of the baryon octet. The only difference between them lying
effectively in the evaluation of the magnetic moment of the 
$\Lambda$-hyperon. 
It may have deeper meaning as $\Lambda$-hyperon being composed of all
different quarks is characterized by zero values of isotopic
spin and hypercharge. 
Without more elaborated theoretical calculations 
of either instanton ball interactions or profile function of 
Nambu-Jona-Lasinio model it is difficult to argue for either of these
models. 

The most important result that can be deduced from \cite{Iwao} and \cite{Kim} 
is that the unitary symmetry model with some 
kind of middle strong interaction and meson cloud contribution lies 
in the base of both descriptions of the magnetic moments of the baryons. 

\section*{ACKNOWLEDGMENTS}

The author wishes to express her hearty thanks to Profs. V.M.~Dubovik
and V.S.~Zamiralov for useful discussions.


\end{document}